\documentclass[superscriptaddress, reprint, amsmath,amssymb, aps,]{revtex4-1}
\usepackage{graphicx}
\usepackage{dcolumn}
\usepackage{bm}
\begin{document}

\title{Self Assembled Clusters of Spheres Related to Spherical Codes}

\author{Carolyn L. Phillips}
\affiliation{Applied Physics Program, University of Michigan, Ann Arbor, Michigan, 48109, USA}
\author{Eric Jankowski}
\affiliation{Department of Chemical Engineering, University of Michigan, Ann Arbor, Michigan, 48109, USA}
\author{Michelle Marval}
\affiliation{Department of Materials Science and Engineering, University of Michigan, Ann Arbor, Michigan, 48109, USA}
\author{Sharon C. Glotzer}
\email{corresponding author \emph{E-mail address:} sglotzer@umich.edu}
\affiliation{Department of Chemical Engineering, University of Michigan, Ann Arbor, Michigan, 48109, USA}
\affiliation{Department of Materials Science and Engineering, University of Michigan, Ann Arbor, Michigan, 48109, USA}
\affiliation{Applied Physics Program, University of Michigan, Ann Arbor, Michigan, 48109, USA}

\date{\today}

\begin{abstract}
We consider the thermodynamically driven self-assembly of spheres onto the surface of a central sphere.  This assembly process forms self-limiting, or terminal, anisotropic clusters ($N$-clusters) with well defined structures.  We use Brownian dynamics to model the assembly of $N$-clusters varying in size from two to twelve outer spheres, and free energy calculations to predict the expected cluster sizes and shapes as a function of temperature and inner particle diameter.  We show that the arrangements of outer spheres at finite temperatures are related to spherical codes, an ideal mathematical sequence of points corresponding to densest possible sphere packings.  We demonstrate that temperature and the ratio of the diameters of the inner and outer spheres dictate cluster morphology and dynamics.  We find that some $N$-clusters exhibit collective particle rearrangements, and these collective modes are unique to a given cluster size $N$.  We present a surprising result for the equilibrium structure of a $5$-cluster, which prefers an asymmetric square pyramid arrangement over a more symmetric arrangement.  Our results suggest a promising way to assemble anisotropic building blocks from constituent colloidal spheres.
\end{abstract}

\maketitle

\section{Introduction}

Anisotropic particles are compelling building blocks for self-assembled materials because their directional interactions can be exploited to create complicated and useful patterns \cite{glotzer07, pine, jones,fejer1, fejer2, shell, jankowski11}.  One way to create anisotropic building blocks is to self-assemble them from simpler particles, where the building block represents a free-energy minimizing structure.  Recently a number of papers have been published synthesizing and simulating compound building blocks that are clusters of spheres\cite{Wales, Sciortino10, Sciortino2004, shell, Ting, Ting2, Manoharan25072003, Lauga, Cho2005, ChoManoharanPine, Cho, Granick}.  Colloidal spheres are attractive candidates for assembly because they can be made from a wide variety of polymers and metals, and their interaction potentials can be tuned with organic ligands, solvents, and salts.  

Here we consider a class of self-limiting, or ``terminal'', colloidal clusters created by self-assembling a small population of one type of particle, the ``halo'' particle (HP), around a second type of particle, the central particle (CP).  The clusters are terminal because the only attractive interaction is between the HP and CP, which are dilute in the fluid of HP, and therefore steric restrictions among co-adsorbed HPs inhibit further growth.   The resulting clusters have structures determined by the interactions among the adsorbed HPs, which self-organize around the CP to minimize their free energy.   

\begin{figure*}[h!t]
\includegraphics[scale=1.0]{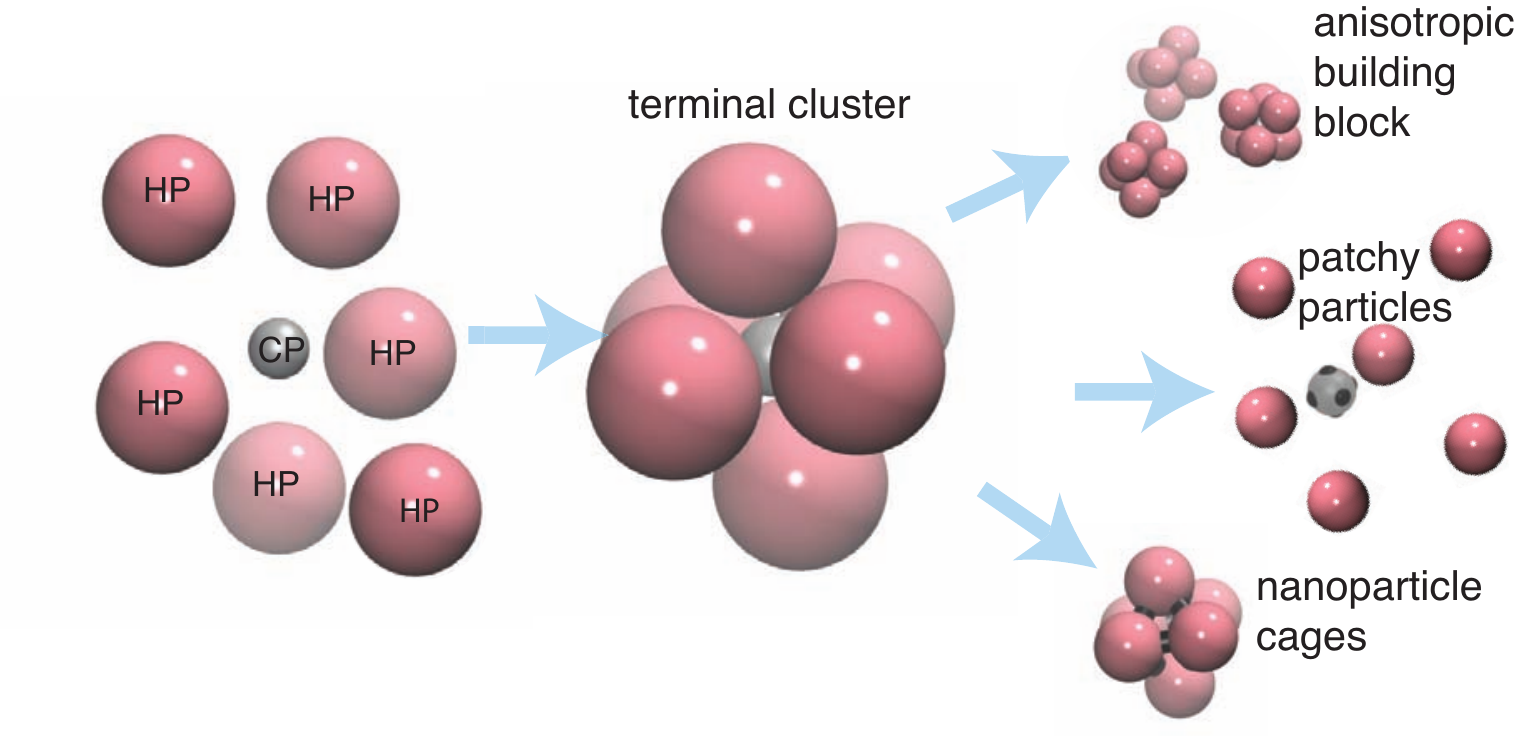}
\caption{(Color) A terminal $N$-cluster with an octahedral structure ($N$ = 6) is self-assembled from a bath of HP and a CP.   This cluster has applications as an anisotropic building block, could be used to manufacture a ``patchy particle''  by imparting patches on the CP at the contact points, or could be locked into a nanocolloidal cage structure.}
\label{makingpatchyparticles}
\end{figure*}

Arrangements of HPs on the surface of a CP have been studied extensively by mathematicians in the context of optimal arrangements of points on a sphere\cite{Toth, njas,extremal,1952}.  The solutions provide a library of anisotropic clusters that can in principle be created with properly designed interactions among the constituent particles.  In this work we study hard sphere HPs that are attractive only to dilute CPs and not to other HPs, thereby producing clusters of HPs around a single CP. The arrangements of these HPs bear comparison to a particular set of solutions, the \emph{spherical codes}, for certain ratios of particle diameters.  We investigate the self-assembly of these clusters as a function of temperature, where entropy controls the equilibrium structure of the cluster, and in a semi-open system, where HP are free to bind and unbind from the CP surface.  We also consider the effect of temperature on the cluster structures and dynamics at deviations from the perfectly dense  packings that correspond to solutions of the spherical code.

This paper is organized as follows.  In Section II we briefly review sphere surface extremal point problems.  In Section III, we introduce the methods we use to study the terminal $N$-clusters, including Brownian dynamics simulations, free energy calculations, and metrics for cluster structure and mobility.  In Section IV, we report the results of our simulations, free energy calculations, and analyses.  We find that terminal $N$-clusters self assemble across a range of diameters and temperatures and the structure of these clusters resemble spherical code solutions. These findings are supported by free energy calculations, which predict cluster sizes and distributions.  Using Brownian dynamics and free energy calculations, we explain the surprising observation of a dominant low-symmetry $N$ = 5 cluster, a deviation from the spherical code prediction.   We calculate changes in cluster structure across a range of diameter ratios and investigate the dynamics for different cluster sizes, including collective modes.  We find that the dynamics for clusters of different sizes are different. In Section V, we discuss  several ways this work can be extended to create more types of anisotropic particles via tuning of the particle interactions, constructing additional shells of particles, and creating structurally reconfigurable particles.  In Section VI we conclude with a summary of our findings. 

\section{Sphere Surface Extremal Points and Spherical Codes}

\begin{figure}[h!]
\includegraphics[scale=0.9]{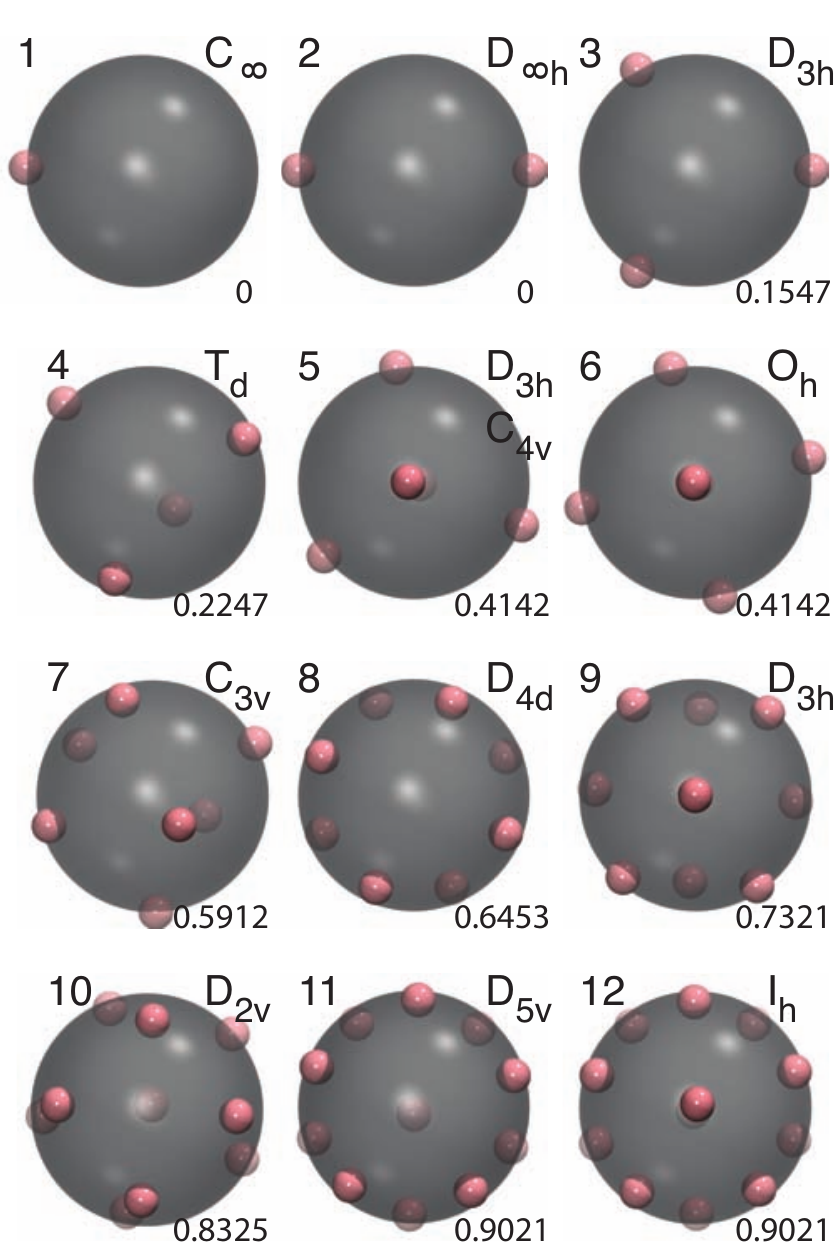}
\caption{(Color) The arrangement of points (pink) that correspond to each spherical code solution for $1\leq N\leq12$.  The point group of each arrangement is shown to the upper right of each arrangement, and the densest packing diameter ratio $D_c/D_h=\Lambda_N$ is shown to the lower right.   For $N=5$, the triangular bipyramid configuration is shown.  Other $N=5$ configurations are shown and discussed in Figures~\ref{fig:SPandTBP}-\ref{fig:SBPDist} .\label{fig:sphericalcodes} }
\end{figure}

The problem of finding extremal points obtained by optimally distributing points on the surface of a sphere to minimize a function $f$ has been well studied in the field of mathematics\cite{1952, edmundson, extremal}.  The problem is typically posed as follows:

 \emph{Given N points on the surface of a sphere of radius R, what arrangement of the N points minimizes a function f}?  
 
 If $f = k\sum^N_{i \neq j} r_{i,j}^{-n}$, where $r_{i,j}$ is the Euclidean distance between the points $i$ and $j$, and $n=1$, minimizing $f$ corresponds to the Thomson problem, whose solution describes the distribution of identical point charges on the surface of a sphere.  As $n \rightarrow \infty$, the problem corresponds to the \emph{spherical code}, (also known as the Fejes T\'{o}th, or Tammes problem), whose solution maximizes the minimum distance between any two sets of points\cite{Toth, njas,extremal,1952}.  Other possible choices for $f$ include minimizing the maximum distance of any point to its closest neighbor, also known as the \emph{sphere covering problem}, and  maximizing the volume of the convex hull of the points.  For each of these problems solutions are exactly known for some values of $N$, while various numerical searches have suggested best solutions for other $N$.  For the functions mentioned, tables of putative solutions up to at least $N=130$ can be found in Ref.~\cite{njas}. 
 
Fig.~\ref{fig:sphericalcodes} depicts the spherical code solutions for $1\leq N \leq12$.  The arrangement of points for ${N}=4$ corresponds to the vertices of a regular tetrahedron, ${N}=6$ an octahedron, ${N}=8$ a square anti-prism, and ${N}=12$ an icosahedron.  The point arrangement of ${N}=11$ is equal to the ${N}=12$ solution minus a single point, or an icosahedron with one truncated pentagonal face.  For each ${N}$, the point group -- the group of isometries that keeps one point fixed -- of the arrangement\cite{extremal} is shown in the upper right corner.  Each optimal arrangement of ${N}$ points on the surface of the sphere is unique except for ${N}=5$ which has a continuum of solutions ranging from a triangular bipyramid (point group $D_{3h}$, shown in Fig.~\ref{fig:sphericalcodes} and Fig.~\ref{fig:SPandTBP}b) to a square pyramid  (point group $C_{4v}$, shown in Fig.~\ref{fig:SPandTBP}a).  All solutions in the continuum have two points at opposite poles of the central sphere and differ by the positions of the three remaining points on the equator.  The square pyramid arrangement is equal to the $N$ = 6 solution minus a single point.  We discuss these structures in detail in Section~\ref{breaking}.  

If the $N$ points represent sphere centers, the spherical code solution corresponds to the densest packing of $N$ hard halo spheres that all ``kiss'' a central sphere.  For any packing of spheres around a central sphere, we define $\Lambda$ to be the ratio of the central sphere diameter, $D_c$, to the halo sphere diameter, $D_h$.  We denote the minimal possible diameter ratio for $N$ spheres, which corresponds to the spherical code solution, as $\Lambda_N$. In Fig.~\ref{fig:sphericalcodes}, $\Lambda_N$ of each arrangement is shown to four significant digits in the bottom right corner.  Notably, $\Lambda_{N=5} = \Lambda_{N=6}$ and $\Lambda_{N=11} = \Lambda_{N=12}$.   In one of mathematics' most famous debates, Isaac Newton and David Gregory argued whether the kissing number of unit spheres ($\Lambda = 1$) is 12 or 13.  Had it been known that a central unit sphere can only be kissed by 13 spheres if their radii is $r\leq 0.9165$, or $\Lambda_{13} = 1.0911$\cite{njas}, this would have settled the question.  Isaac Newton's conjecture that the kissing number is 12 was not proven until 1953\cite{Aste}.

We also note that the spherical code solutions for $N$ = 3-12, except $N$ = 5, are rigid or jammed.  They contain no ``rattlers'', defined as spheres not in isostatic contact with other spheres\cite{rigiditysc,Tarnai}, and cannot be deformed other than global isometries\cite{rigiditysc,Tarnai}.

\section{Methods}
To predict and compare the terminal $N$-clusters of halo particles bonded to central particles we use computational tools that sample equilibrium statistical mechanical ensembles.  In particular, Brownian dynamics simulations of model particles are used to perform computer experiments of self-assembly and the results of these simulations are compared against cluster probabilities calculated from a free energy analysis based upon numerical partition function calculations\cite{jankowski11}.  We also calculate detailed structural and dynamic quantities for each cluster.

\subsection{Hard Sphere and Sticky Sphere Model} \label{spheremodel}
\begin{figure}[h]
\includegraphics[scale = 1.0]{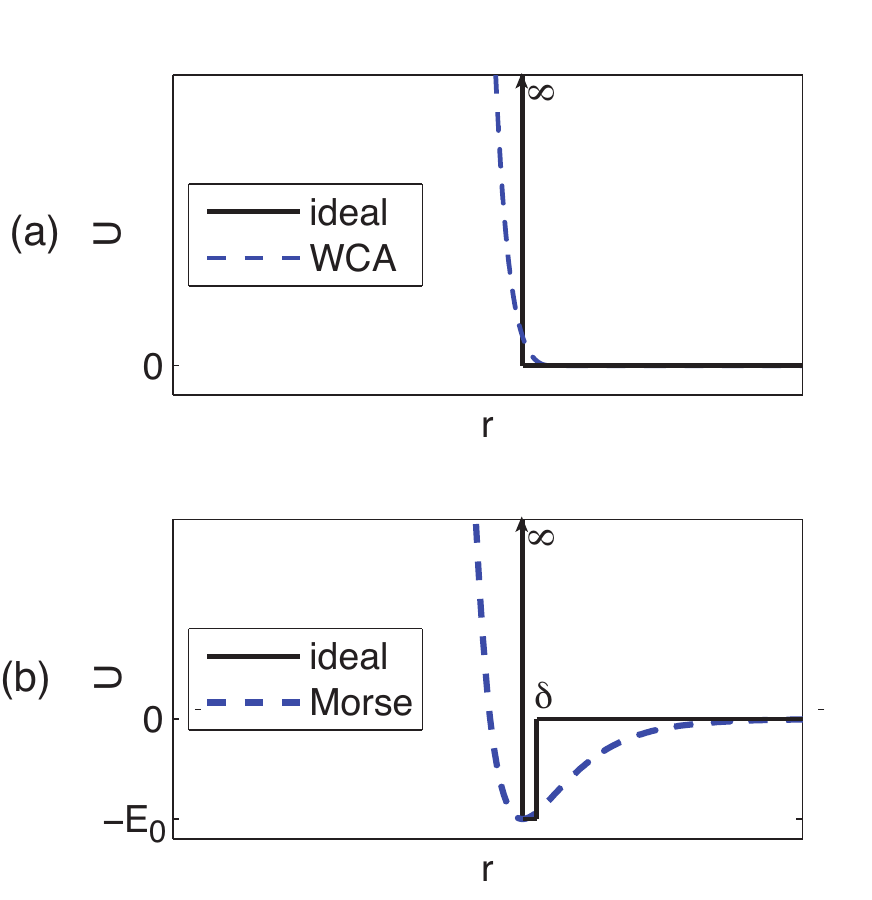}
\caption{(a) A mathematically ideal hard particle interaction is shown in solid black compared to the hard particle interaction (in dashed blue) given by the WCA potential (Eqn.~\ref{eqn:wca}). (b) A sticky sphere with a kissing contact potential when $\delta \rightarrow 0$ is compared to a model sticky sphere (in dashed blue) given by the Morse potential (Eqn.~\ref{eqn:morse}).  \label{fig:idealmodel} }
\end{figure}

In a semi-open system, the spherical code solutions of Section II correspond to perfectly hard spheres adsorbed on a perfectly sticky sphere.  Mathematically, perfectly hard spheres are points interacting via a function that steps from infinity to zero (Fig. \ref{fig:idealmodel}a) and perfectly sticky spheres are points interacting via the same function plus an infinitely narrow square well function (Fig. \ref{fig:idealmodel}b).  
In this work we use radially-shifted Weeks-Chandler-Andersen (WCA) and Morse models of hard and sticky spheres, respectively, which allow for computational efficiency as well as direct comparison with their ideal mathematical counterparts (Fig \ref{fig:idealmodel}).  They also capture, in a general sense, the repulsive and attractive interactions of the constituent particles we have in mind. As nanoparticle synthesis continues to mature, the types of interactions that can be used to guide the self-assembly of small particles can be precisely tuned over wide ranges of length and energy scales, and the models used in simulations can be suitably adjusted. 

The radially-shifted WCA potential is given by\cite{weeks:5237}  

\begin{eqnarray}
U_{WCA} = \left\{ 
\begin{array}{l l}
  4\epsilon \left(\left(\frac{\sigma}{r-\alpha}\right)^{12}-\left(\frac{\sigma}{r-\alpha}\right)^{6}\right)&+\epsilon \\  & r < r_{cutoff}\\
  0  & r \geq r_{cutoff}\\ \end{array}. \right. \label{eqn:wca}
  \end{eqnarray}
The shifting parameter $\alpha$ is defined as $\alpha=\sigma_h-\sigma$, where $\sigma_h$ is the WCA ``diameter'' of the HP, and $r_{cutoff}=2^{1/6}\sigma+\alpha$.  The interaction between two HPs can be made arbitrarily hard relative to their size by increasing $\sigma_h$.   The cost of increasing $\sigma_h$ is that the dimensionless time $\tau$ that elapses over each time step is reduced as $\tau \propto1/\sigma_h$.  We choose $\sigma_h = 3\sigma$ for its computational efficiency and its relatively ``hard'' modeling of spheres.  The energy parameter $\epsilon$ also determines the ``hardness'' of the HP, as a larger energetic penalty to overlapping corresponds to a ``harder'' potential.  The cost of increasing $\epsilon$ is that a smaller simulation time step is needed to model a steeper function. The energy parameter is set to $\epsilon = (0.1/T)$, where $T$ is the temperature of the simulation, so that the hardness of the HP is independent of temperature.  

Using Eqn.~\ref{eqn:wca} to model hard HPs is effective because of the large potential energy penalty associated with two HPs approaching closer than $3\sigma$.  However, due to the soft nature of the potential, spheres with sufficient kinetic energy can, in principle, approach as close as $2\sigma$.  It is therefore useful to determine the effective hard particle diameter of HP modeled by Eqn.~\ref{eqn:wca}. We use the Barker-Henderson equation\cite{PT} to calculate the effective diameter
\begin{equation}
D_{h,e} = \int_0^{\sigma_{BH}} (1-e^{-\beta u(r)})dr \simeq 3.0786\sigma \label{eqn:deff}
\end{equation}
where  $\beta = 1/k_BT$, $u(r) = U_{WCA}$ and the potential is zero at $\sigma_{BH} = 2^{1/6}\sigma + \alpha$.
For the purpose of assessing the error in our calculations based on the effective diameter, we can characterize two HPs as contacting when the interaction energy between them is in the range $0 < U_{WCA} < 10k_BT$, which corresponds to $3.0\sigma < D_{h,e} < 3.1225\sigma$.  

The radially-shifted Morse potential\cite{morse} used to model the ``kissing'' contact potential between the HP and CP is given by 
\begin{eqnarray}
U_{M} = \left\{ 
\begin{array}{l l}
  E_0\left(e^{-2\beta(r-r_0)} -2 e^{-\beta(r-r_0)} \right) & \quad r < r_{cutoff}\\
  0 & \quad r \geq r_{cutoff}\\ \end{array} \right. \label{eqn:morse}
  \end{eqnarray} 		 
where $E_0=5$ determines the depth of the energy well, $\beta = 5.0/\sigma$ determines the width of the energy well, and $r_0$ determines the radial displacement of the bottom of the energy well.  The Morse potential interaction range is truncated at $r_{cutoff}=2.5 \sigma+ (r_0-\sigma)$.  

An effective CP diameter can be calculated by defining an HP and CP as \emph{bonded} when the distance between the two particle centers is at the minimum of Eqn.~\ref{eqn:morse}.  More properly, an HP and CP are bonded when they remain positionally correlated because the HP remains within a given displacement of the CP.  We use the minima of Eqn.~\ref{eqn:morse} and the effective diameter of Eqn.~\ref{eqn:deff} to define an effective CP diameter $D_{c,e} = 2r_0 -D_{h,e}$.  The ratio of the CP to HP diameter is therefore $\Lambda^{m}$ =  $D_{c,e}/D_{h,e}$, (\emph{m} for \emph{molecular dynamics}).  We choose to keep the hardness of the HP-HP interaction constant for all the MD simulations by holding $D_{h,e}$ (i.e. $\sigma_h$) constant while varying $r_0$ to change $D_{c,e}$.  

The non-infinitesimal potential well width of Eqn.~\ref{eqn:morse} permits the bond between the CP and HPs to stretch a small amount while remaining bonded.  At some $\Lambda^m$ ratios, this stretching, though small, may be enough to accommodate an additional HP bond to the CP.  It is therefore useful to define the CP \emph{bond-stretched effective diameter} $D_{bs,c,e} = 2\bar R_{hc}-D_{h,e}$, where $\bar R_{hc}$ is the average center-to-center distance between a bonded HP and CP measured in a simulation.  The bond-stretched diameter ratio is defined as $\Lambda^{bs} = D_{bs,c,e}/D_{h,e}$ ($bs$ for \emph{bond-stretched}).  $\Lambda^{bs}$ is always greater than $\Lambda^{m}$.  When the cluster is loosely packed, the difference between the two measures converges to zero. 
 
\subsection{Brownian Dynamics}
 
To model mixtures of halo particles and central particles assembling in a thermal bath we perform Brownian dynamics (BD) simulations, implemented in HOOMD-blue\cite{HOOMD}.  The natural units of this system are: the effective diameter of the HP, $D_{h,e} = 3.0786\sigma$; the mass of a HP, $m$; and the depth of the HP-CP energy well, $E_0$.  The volume fraction, $\phi$, is defined as the ratio of the total volume of the HPs and CPs to the simulation box volume, the dimensionless time is $t^*=t/(D_{h,e}\sqrt{m/E_0})$ , and the dimensionless temperature is $T^*=k_BT/E_0$.  We use periodic boundary conditions. Each particle is subjected to conservative, random, and drag forces, and its motion is governed by the Langevin equation discussed further in \cite{Zhang,Iac2005,Iac2008}.  We use a value for the drag coefficient  $\gamma = 0.726$ $m/t^*$.  The same drag coefficient is applied to HPs in both the free and bound state.  The conserved forces between particles are per Eqns.~\ref{eqn:wca} and \ref{eqn:morse} above.  

\subsection{Free Energy Calculations}\label{sec:fec}
The relative probability of finding a particular cluster of $N$ HPs bound to a CP can be predicted using free energy calculations detailed in references \cite{McGinty, Meng, jankowski11}.  For a given $\Lambda$, the partition function is defined by the appropriately weighted sum over all possible  configurations of $N$ HPs bound to a CP for $N = 1$ to $\infty$. The contributions of the distinguishable microstates to the partition function are calculated numerically.  

The partition function is calculated assuming ideal hard spheres and sticky spheres (Fig.~\ref{fig:idealmodel}).  Given HPs of diameter $D_h$ and a CP of diameter $D_c$, the interaction potential between ideal HPs is defined as,
\begin{eqnarray}
U_{H-H}(r) = \left\{ 
\begin{array}{l l}
  \infty & \quad r < D_h\\
  0 & \quad r \geq D_h\\ \end{array} \right . 
  \end{eqnarray}

\noindent and the interaction potential between an ideal HP and an ideal CP is defined as 
\begin{eqnarray}
U_{H-C}(r) = \left\{ 
\begin{array}{l l}
  \infty & \quad r < (D_c + D_h)/2\\
  -E_0 & \quad r = (D_c + D_h)/2\\
  0 & \quad r >  (D_c + D_h)/2\\ \end{array} \right. 
 \end{eqnarray}

We define $\Lambda^f =  D_c/D_h$ (\emph{f} for \emph{free energy calculation}).  As in section \ref{spheremodel}, to vary $\Lambda^f$, $D_h$ is held constant (set to $D_{h,e}$ from Eqn.~\ref{eqn:deff}) and $D_c$ is changed.  

If $\Lambda_{N=M} \leq \Lambda^f < \Lambda_{N=M+1}$, then configurations of ${M}$ HPs bonded to the CP minimize the potential energy and configurations with more than $M$ HPs have infinite potential energy (zero probability).  Configurations with fewer than $M$ HPs bonded to the CP increase the entropy of the cluster.  When $k_BT\approx E_0$, the free energy can be minimized by clusters with fewer than $M$ HPs, because the entropy gained by the remaining HPs on the CP balances the increase in potential energy.  In the grand canonical ensemble, at a fixed $\Lambda^f$, the probability of observing a particular cluster $s$ is given by the Boltzmann distribution:
\begin{equation}
P_s = e^{-\beta F_s} =\frac{\Omega_s  e^{-\beta (U_s - \mu N)}}{\mathcal{ Z}}
\end{equation}
where $\mathcal{Z}=\sum_s \Omega_s \exp (-\beta(U_s - \mu N))$ is the partition function and $U_s - \mu N =  NE_0$.  Without loss of generality, we treat $\mu$ =0.  (A non-zero $\mu$ will only induce a uniform temperature shift in our final results.)

In practice, calculating $\mathcal{Z}$ exactly is difficult, but by assuming that only a small number of clusters contribute to $\mathcal{Z}$\cite{McGinty,Meng,jankowski11}, the relative probabilities of these clusters can be determined.  As in \cite{McGinty,Meng,jankowski11}, the degeneracy $\Omega_s $ can be written as a product of three independent terms, the translational, $Z_t$, rotational, $Z_r$, and vibrational $Z_v$ partition functions.  The translational partition function is approximately equal for all the clusters because they are all small compared to the accessible volume, and thus contributes equally to the $\Omega_s$ of each cluster.  

To calculate the rotational and vibrational partition functions for an $N$-cluster, we first assume an equilibrium configuration defined by $N$ HPs in a spherical code configuration at a radial displacement of ($D_c$ + $D_h$)/2 from the CP.  The rotational partition function is then calculated as $Z_r = c_r \frac{\sqrt{I}}{\kappa}$, where $c_r$ is a temperature-dependent constant that is the same for all the clusters, $I$ is the determinant of the moment of inertia tensor, and $\kappa$ is the symmetry number of the spherical code configuration under rotation.  Each sphere is given a unit mass.  The vibrational partition function is proportional to the product of the vibrational freedom, or freedom to rattle, of each sphere in the cluster.  The vibrational freedom of each HP can be measured as the fractional area of the surface of the CP it has access to, subject to the restrictions imposed by its neighboring spheres. We approximate the vibrational area available to a given HP in a particular configuration by using a Monte Carlo numerical approach whereby new positions for the HP are randomly generated and accepted if the HP does not overlap another HP.  The accessible vibrational area is proportional to the total number of accepted positions that are part of a contiguous area that includes the HP's original position divided by the total number of random trials.   If $\Lambda^f = \Lambda_N$, when the diameter ratio matches the spherical code ratio, then most, if not all, of the spheres in an $N$-cluster are jammed and have no vibrational freedom.  

The free energy calculation is approximate, as it does not consider the contribution of collective modes of HP motion to $\mathcal{Z}$, which, in certain systems can help stabilize one configuration over another \cite{haji-akbari:194101}.  As we show in Section \ref{sec:struct}, each cluster has a small $\Lambda$ range, $\Lambda > \Lambda_N$ where collective modes are not present, and only local rattling is observed.  Outside this range, we expect some error in the calculation of relative probabilities to accumulate.  The benefit of this free energy approximation is demonstrated by both its favorable comparison to predictions made by BD simulations and by its ability to rapidly predict the entire phase diagram.  Applying a more computationally intensive method to perform an exact free energy comparison would be an interesting topic for future study.

\subsection{Structure and mobility measures of a cluster} \label{sec:constrained}
The HPs in an $N$-cluster for $\Lambda=\Lambda_N$ are confined to a unique $N$ spherical code solution and cannot rearrange or even rattle for $N$ = 3 to 12, excepting $N$ = 5.  For $\Lambda >> \Lambda_N$, the HPs are free to randomly arrange on the surface of the CP.  We aim to understand the structure and dynamics of the $N$-cluster between these extremes.  We perform BD simulations of pre-assembled clusters wherein the HPs are restricted to the surface of the CPs, a constraint imposed during the integration of the equations of motion.  This allows the dynamics of HP rearrangement to be isolated from the dynamics of assembly and disassembly and prevents any stretching of bonds from influencing the structures observed.  Similar to section \ref{spheremodel} the CP diameter is defined as $D_{c,e} = 2r_0-D_{h,e}$, where $r_0$ is the fixed distance between the CP and HP centers and $D_{h,e}$ is the same as defined in Eqn.~\ref{eqn:deff}.  The CP to HP diameter ratio in the constrained system is thus $\Lambda^c =D_{c,e}/D_{h,e}$ ($c$ for \emph{constrained}) and varied by changing $r_0$.  $\Lambda_c$ is initialized such that the HP can be sparsely randomly distributed on the surface (i.e. $\Lambda^c>>\Lambda_N$), and then slowly decreased over the course of a simulation to a target $\Lambda^c$.

The angular displacement between two HP bound to the same CP, or $\theta=\angle ACB \;$  is defined by the centers of two HPs $A$ and $B$ and the CP $C$.  To characterize the structure of the cluster, the distribution of angular displacements between pairs of HPs, $n(\theta)$ for $\theta = [0,\pi]$ are measured for a fixed $N$ and $\Lambda^c$ over all HP pairs every $10^4$ time steps during a simulation with $10^9$ total time steps.  The value of $n(\theta)$ for a given $N$ and $\Lambda^c$ represents the likelihood of finding an HP at an angle $\theta$ relative to a given HP, and $\int n(\theta)d\theta=N-1$.  $n(\theta)$ is analogous to a pair correlation function.

To characterize the dynamics we calculate the time scale over which $\theta$ is no longer correlated with itself.  We define the mobility parameter $\tau$ from 
\begin{equation}
C(\theta(t),\theta(t+\delta t))=e^{-\tau t}
\end{equation}
where $C(\theta(t),\theta(t+\delta t))$ is the normalized angular autocorrelation function and $t$ is time.  In this work,  $\tau$ has units of 1/10,000 time steps.  The more mobile an HP is on the surface of the CP, the more rapidly its angular displacement with respect to other HP decorrelates.   When the rate of decay of angular correlations is zero, all the HP in the cluster are fully caged.  We only calculate $\tau$ for clusters that display more than one distinct peak in their $n(\theta)$ distributions so that position swapping can be distinguished from local rattling.  The lower bound on the $\tau$ measurement is $1.5\cdot10^{-4}$ because below this the HP position swaps occur too infrequently over a $10^9$ time step simulation for accurate values of $\tau$ to be measured.   We calculate $\tau$ as a function of $N$ and $\Delta \Lambda^c = \Lambda^c - \Lambda_N$, or the difference between the diameter ratio of the $N$-cluster and the $N$ spherical code solution ratio.   Because the HPs in the simulation are not perfectly hard and are constrained to the CP surface, it is possible for meaningful measurements to be made when $\Delta \Lambda^c < 0$.

\subsection{The calculation of $\Lambda$}
In this paper, to elucidate different properties of $N$-clusters, several calculation methods are used, necessitating four different ways to determine $\Lambda=D_c/D_h$, the ratio of the central particle diameter, $D_c$, to halo particle diameter, $D_h$.  Each way was chosen to best represent the effective diameters of the HP and CP in the particular method. These $\Lambda s$ are comparable to each other and to the spherical code solution ratios, $\Lambda_N$ of Fig.~\ref{fig:sphericalcodes}.  We indicate the calculate type by the superscript $x$ of $\Lambda^x$, where $x \in \{m, bs, c, f\}$, where the $m$, Brownian (molecular) dynamics; $bs$, bond-stretched; $c$, constrained; and $f$, free energy calculation, as defined above. 

\section{Results}

\begin{figure*}[htp]
\includegraphics{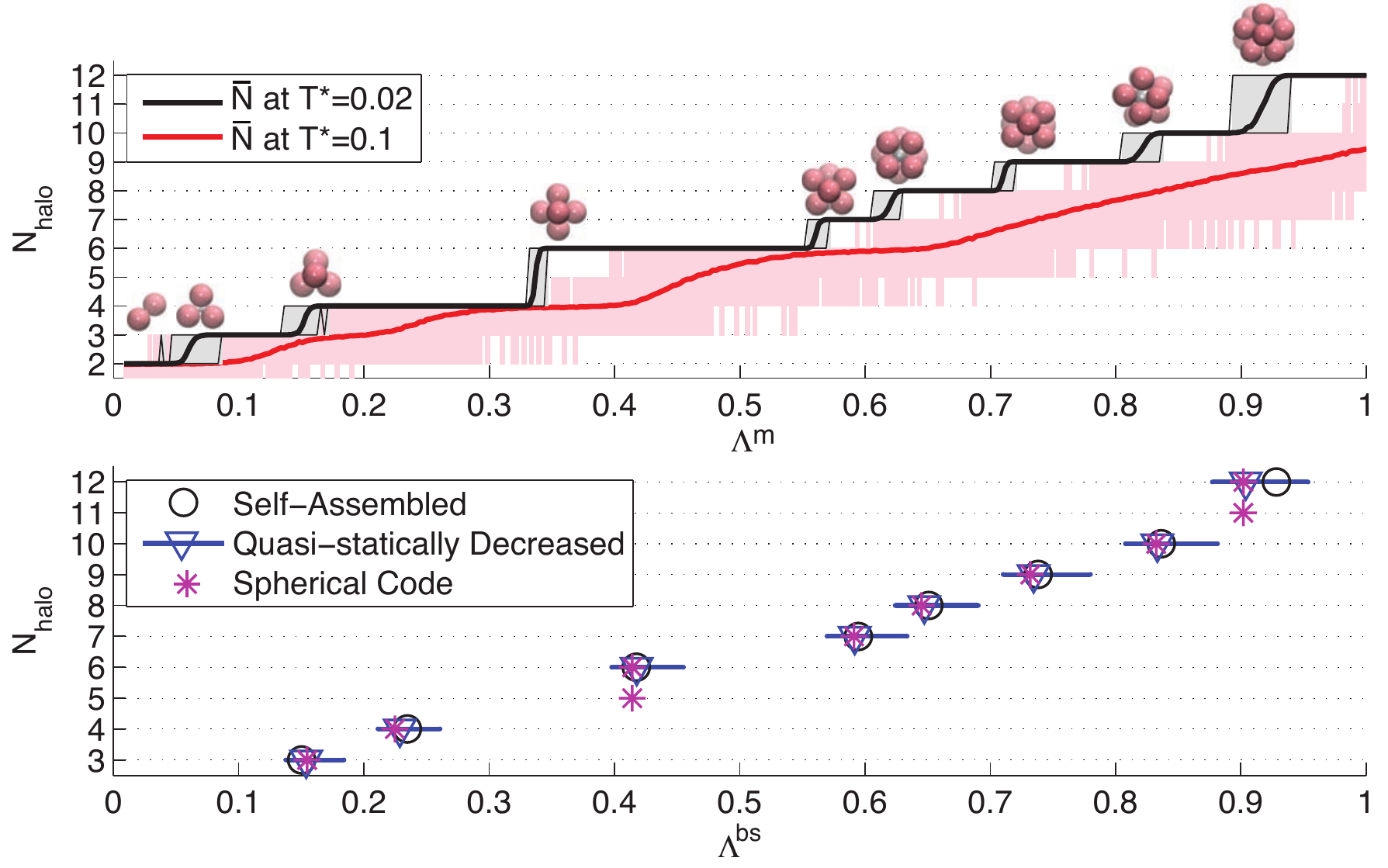}
\caption{(Color) Top: The $N$ clusters that self-assemble as a function of $\Lambda^m$ and temperature is shown.  The average $N$ of the self-assembled cluster at $T^* = 0.02$ is shown as a black line.  The maximum and minimum N in the simulation is shaded grey.  The average $N$ of the self-assembled cluster at $T^*=0.1$ is a red solid line.  The maximum and minimum N in the simulation is shaded pink. Example clusters self-assembled in simulation are shown.  Bottom: Accounting for bond-stretching and the effective diameter of the HP, the lowest ratio where a cluster of size $N$ observed in the quasi-statically decreasing simulation (blue triangles) and for the self-assembled simulations (black circles) are compared to the spherical code predictions (pink star).  Error bars for the quasi-static simulation ratios are generated from the contact range of two HP.\label{fig:avha} }
\end{figure*}

\begin{figure*}[htp]
\includegraphics[scale = 1.0]{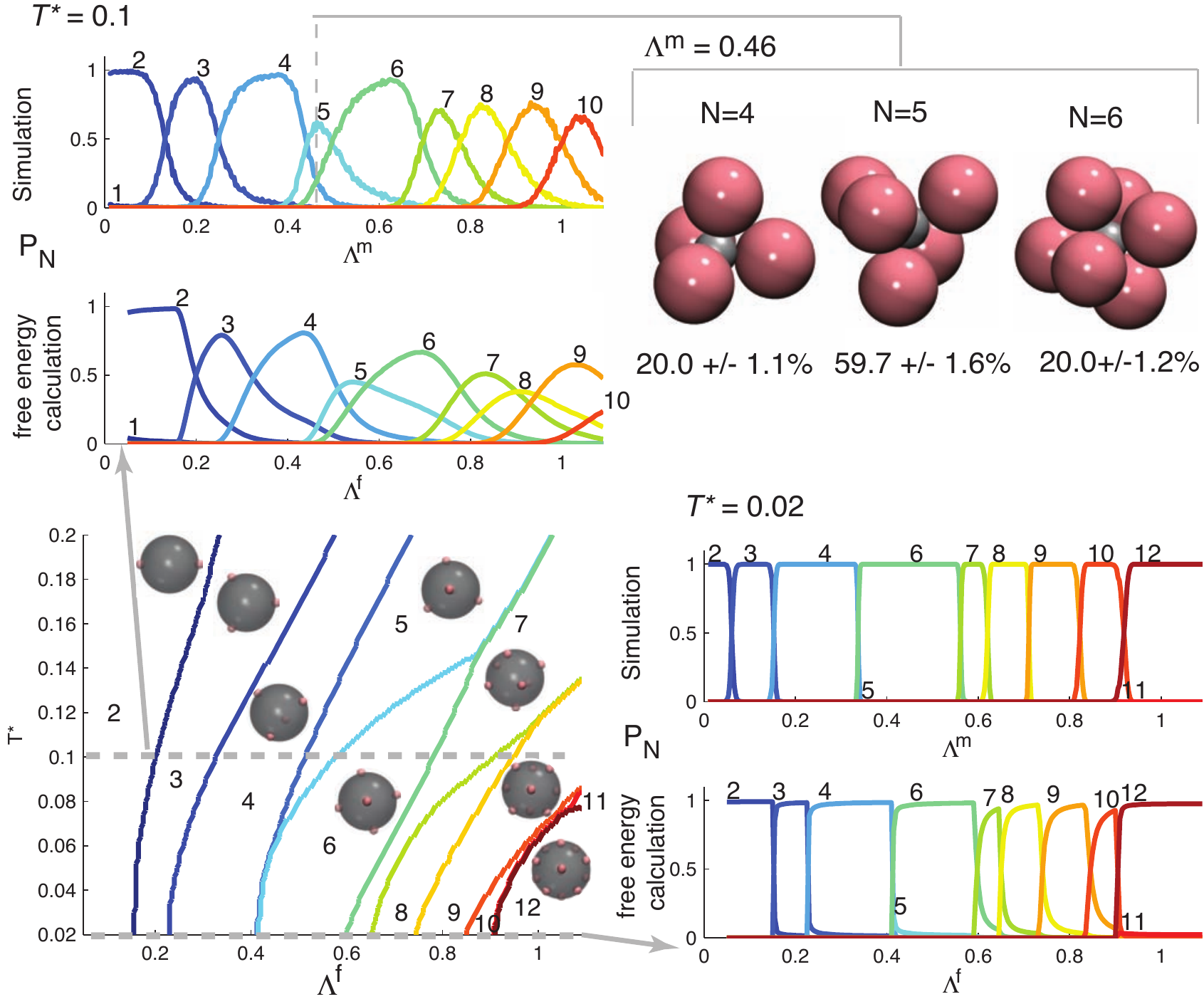}
\caption{(Color) The distributions of cluster sizes as a function of temperature and $\Lambda$ as given by the free energy calculation and the BD simulations are compared.   Bottom left corner: phase diagram of the free energy prediction of the most probable cluster size.  Lower right and upper left corners: in-page slices of the probability of finding each cluster size $P_N$ as predicted by the free energy calculation and BD simulation at the high and low temperature.  Upper right corner: the three most common clusters found in the BD simulation at the high temperature and $\Lambda^m = 0.46$.  \label{wholephase}}
\end{figure*}

\subsection{Self-assembly and free energy of $N$-clusters}
Using Brownian dynamics we simulate the self-assembly of clusters as a function $\Lambda^m$ and at two different temperatures to investigate the effect of thermal noise on the distribution of stable terminal $N$-clusters.  We compare these results to the known spherical code solutions and to free energy calculations.

Brownian dynamics simulations of self-assembly are initialized by placing $1000$ CPs on a cubic lattice, spaced so as to behave as independent systems.  The lattice is embedded in a bath of HPs  at a total volume fraction of $\phi = 0.24-0.27$.  The bath contains a minimum of four times as many HPs per CP as the maximum cluster size observed for that $\Lambda^m$.   We perform a total of 760 simulations of 20$\times 10^6$ time steps, with time step size $\Delta t^* = 0.00363$ at low ($T^* = 0.02$) and high ($T^* = 0.1$) temperatures.  With this set of simulations, we calculate the cluster size distribution as a function of $\Lambda^m$.  

At the low temperature we observe that the cluster sizes are highly monodisperse as a function of $\Lambda^m$.  In Fig.~\ref{fig:avha}, the mean cluster size assembled at the low temperature, $T^*=0.02$ for  $0.01<\Lambda^m <1$ is shown as a black solid line.  Grey shading indicates the range of cluster sizes observed at a particular $\Lambda^m$.  Over this range of $\Lambda^m$ clusters are uniform in size except when $\Lambda^m$ is near a value where there is a transition from one mean cluster size to another.  At these transitions, we observe a narrowly distributed mixture of cluster sizes; e.g., at $\Lambda^m =0.71$ for the $T^*=0.02$ curve, we find equal numbers of clusters containing $8$ or $9$ HPs. 

In comparison to the low temperature data, the clusters at high temperature are both smaller on average, and have a broader distribution of sizes as a function of $\Lambda^m$.   In Fig.~\ref{fig:avha}, we show the distribution of clusters assembled at high temperature, $T^*=0.1$ (red). The region shaded pink represents the range of cluster sizes measured at a given $\Lambda^m$ at $T^*=0.1$.   At $\Lambda^m =0.71$ for the $T^*=0.1$ curve, we now observe clusters of 5, 6 7, and 8 HPs.  We also observe that the $N$ = 5 and $N$ = 11 clusters are not stable at any $\Lambda^m$ at low temperature but are present in the broader distribution of clusters at high temperature.
 
To test the stability of the self-assembled clusters at low temperature, we perform a simulation wherein the diameters of the CPs in large pre-assembled clusters are slowly decreased.  A single system with  $\Lambda^m$ = 0.9489 is equilibrated for 20$\times 10^6$ time steps at $T^* = 0.02$, at which time every CP is bonded to 12 HPs.  Subsequently $D_{c,e}$ is decreased at a rate of $4.833\times10^{-8}\sigma/\Delta t$ until  $\Lambda^m$ = 0.0101.  As discussed in reference \cite{Phillips}, this decrease in the diameter is slow enough that the system remains quasi-static, i.e. the system is approximately in equilibrium.  For this system, as $\Lambda^m$ approaches a transition ratio, 1-3 HPs detach from a given CP and re-enter the bath, until only two HP are bonded to each CP.  In effect, this quasi-statically decreased $D_{c,e}$ simulation disassembles the clusters as a function of $\Lambda^m$.  

If bond-stretching is taken into account, we find that at a low temperature ($T^* = 0.02$) the $N$-clusters self-assemble at the $\Lambda$ ratio predicted by the spherical code solutions.   In tightly packed clusters, bond stretching makes $\Lambda^{bs} > \Lambda^m$.
In the bottom plot of Fig.~\ref{fig:avha}, the lowest $\Lambda^{bs}$ at which a cluster of size $N$ is observed for the self-assembled (black circles) and quasi-statically decreased (blue triangles) simulation data is shown and compared to the spherical code $\Lambda_N$ ratio (pink stars).  Blue error bars indicate the $\Lambda^{bs}$ ranges from quasi-statically decreased simulations, generated by assuming that the true diameter of an HP is the limits of the contact range defined in section \ref{spheremodel}.  Good correspondence between the predicted and measured ratios is observed when bond-stretching and the appropriate effective diameters of the particles is accounted for. 

We calculate the free energies of all clusters from $N=2$ to $N=12$ over a temperature range of $0.02\leq T^*\leq0.2$ and diameter ratio range of $0.05\leq \Lambda^f \leq 1.09$.  In the bottom left plot of Fig.~\ref{wholephase}, we report a ``phase diagram'' of the most probable cluster at each combination of $T^*$ and $\Lambda^f$.  The plots in the bottom right and top left of Fig.~\ref{wholephase} show data from an in-page slice of the phase diagram at the low and high temperature, $T^*=0.02$ and $T^*=0.1$, and directly compare it to cluster distributions from the BD simulation data of Fig.~\ref{fig:avha}.  For example, the three clusters and probabilities depicted in the upper right of Fig.~\ref{wholephase} are from a single high temperature BD simulation with $\Lambda^m$ = 0.46. 

We see that the free energy calculations support the findings of the BD simulations.  At high temperature and at a given $\Lambda$, both show a decrease in cluster size relative to the low temperature data, as well as a broadening in the distribution of cluster sizes.  Discrepancies in peak height and shape between the two predictions in Fig.~\ref{wholephase} are likely due to the soft sphere approximation, not accounting for the change in the contact energy or effective diameter due to bond-stretching, and also to neglecting the collective vibrational modes in the free energy calculations.  However, the free energy calculation shows that most of the features of the BD simulations at higher temperatures can be attributed to the offsetting of the increase in potential energy (i.e. fewer bonded HPs) by the commensurate increase in vibrational freedom of the remaining bonded HPs.

Consistent with the BD simulation data, the free energy calculation also predicts that $N$ = 5 clusters are not stable at low ($T^*<0.06$) temperatures.  Spherical code solutions indicate that the densest $N$ = 5 clusters occur at the same $\Lambda$ as the densest $N$ = 6 cluster.  Thus, at low temperature, when the free energy is dominated by the potential energy term, the $N$ = 6 cluster is always stable over an $N$ = 5 cluster.  However, the free energy calculation predicts that at $T^*>0.06$ there is a $\Lambda$ range where an $N$ = 5 cluster is the most probable cluster.  This $\Lambda$ range is observable in the high temperature BD simulation data. The stabilization of the $N$ = 5 cluster over the $N$ = 6 cluster at higher $T^*$ arises from the non-negligible contribution of the vibrational partition function, the only term in the partition function that significantly differs between the two clusters.   The $N$ = 11 cluster is similarly predicted to be unstable at low temperatures but stable over the $N$ = 12 cluster at a higher temperature.  The free energy calculation also predicts an entropic stabilization of the $N$ = 7 and 9 clusters over the $N$ =  8 and 10 clusters, respectively, at higher $T^*$ and ``triple points'', at which the probabilities of three clusters (e.g. 4, 5, and 6; or 7, 8, and 9) are equal. 

\subsection{Structure of $N$-clusters} \label{sec:struct}

\begin{figure*}[thp]
\includegraphics[scale = 1.0]{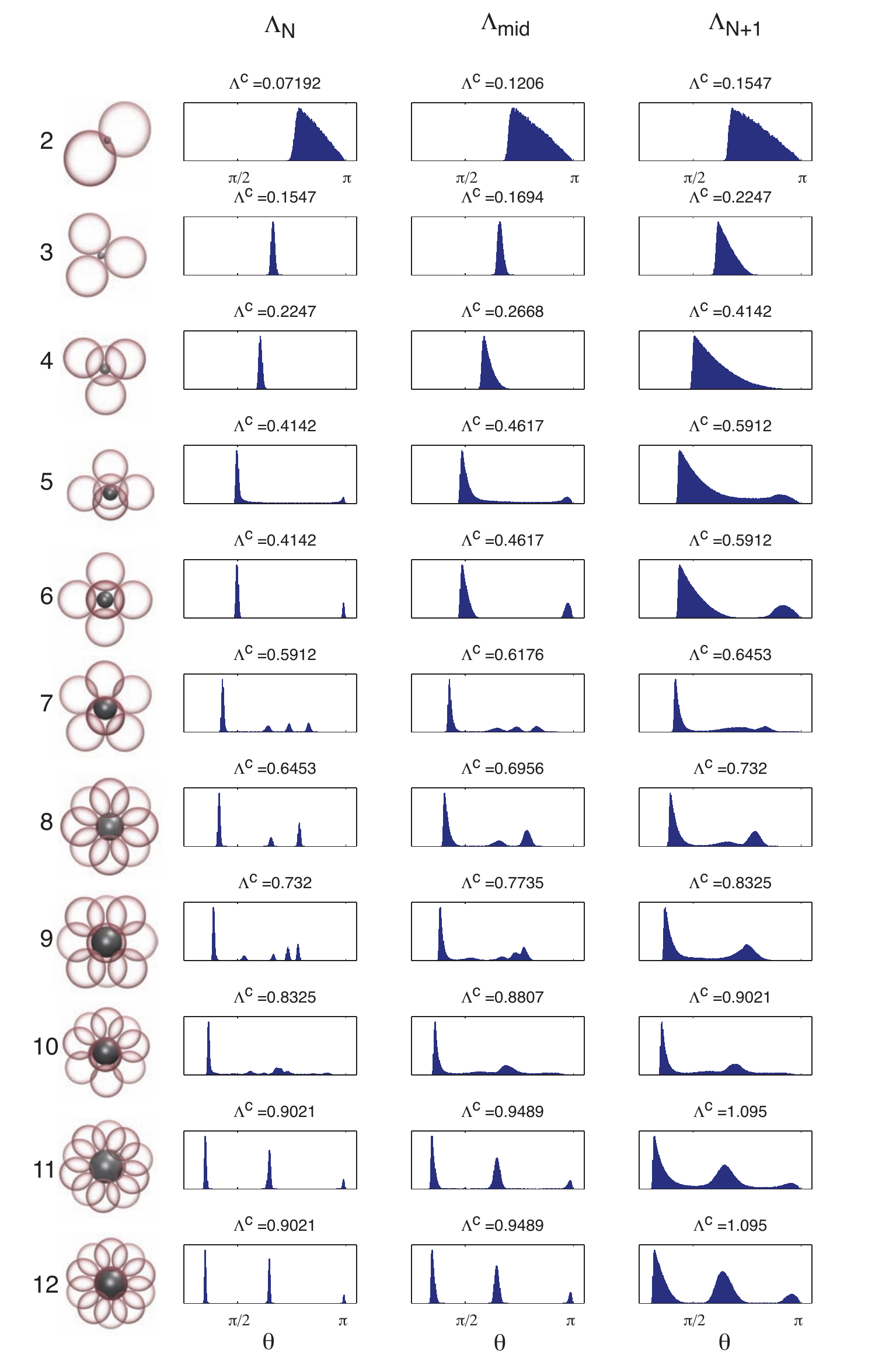}
\caption{(Color) \label{fig:Melting} The distribution of angular displacements $n(\theta)$ for each cluster.  The $n(\theta)$ shows a structural fingerprint particular to each cluster.}
\end{figure*}

\begin{figure}[h!]
\includegraphics[scale = 1.0]{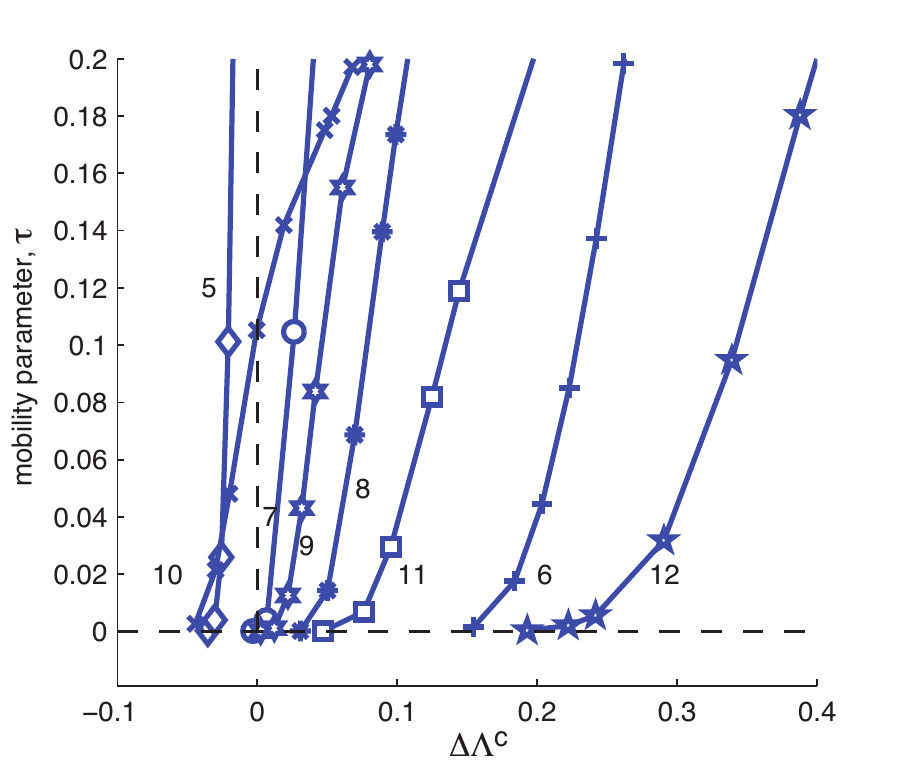}
\caption{\label{fig:mobility} Cluster mobility as a function of the $\Delta \Lambda^c = \Lambda^c - \Lambda_N$.  Note that for the $N$ = 6 and $N$ = 12 clusters, the HPs do not become measurably mobile for $\Delta \Lambda^c >>$ 0.   At the other extreme, $N$ = 5 and $N$ = 10 are mobile for $\Delta \Lambda^c <$ 0.  Each data point is extracted from a linear fitting with a goodness of fit $>$ 0.95.}
\end{figure}

We next consider how the structure of each $N$-cluster changes as $\Lambda > \Lambda_N$.  Clusters that have a large range of $\Lambda$ over which their structure is ordered and stable are desirable targets for synthesis.   We investigate the structure and dynamics of the clusters by modeling HPs constrained to the surface of a CP, as described in Section \ref{sec:constrained} for different $N$ and $\Lambda^c$. 

In simulation, we observe that a cluster of size $N$ generally exhibits three different dynamics over different ranges of $\Lambda^c$.  In the first range, each HP remains locally caged.   Each HP of the $N$-cluster can be assigned to one point of the $N$ spherical code solution of Fig.~\ref{fig:sphericalcodes} and that mapping remains invariant under the dynamics of the cluster.  Like an atom in a crystal lattice, each HP rattles about its point.  In the second range, each HP can almost always be assigned to one point of the spherical code solution, however the mapping does not remain invariant under the dynamics of the cluster.  The HPs sporadically rearrange but are still generally found rattling about the points of the spherical code solution.  In the third range, the HPs cannot be assigned to points of the spherical code solution and move freely on the surface.  Between the second and third range, we suspect that there is no distinct measurable boundary, but simply an increasing likeliness of a cluster being in ``transitional'' states.  Below we show how the two measures introduced in section ~\ref{sec:constrained} capture the signature features of these ranges.

In Figure \ref{fig:Melting}, the distribution of angular displacements, $n(\theta)$, is shown for $2 \leq N \leq 12$ HPs constrained to the surface of a CP at $T^* = 0.02$.   For each $N$, $n(\theta)$, is shown for three different $\Lambda^c$, corresponding to $\Lambda_N$,  $\Lambda_{N+1}$ and a midpoint between the two.  These three distributions are shown in order of increasing $\Lambda^c$ from left to right.  Note that for $N=2$, $\Lambda_{N=2}$ is zero, as it is always possible to add a second HP to a CP with one bound HP, regardless of CP size.  In this case, we arbitrarily choose the minimum $\Lambda^c$ = $\Lambda_{N+1}/2$.  

For each cluster we observe a unique $n(\theta)$ structure fingerprint that softens as $\Lambda^c$ increases.  For $N \leq 4$, each HP has one equidistant ring of neighbors, resulting in $n(\theta)$ having a single peak that broadens as $\Lambda^c$ increases.  For $N > 4$, each $n(\theta)$ has multiple peaks.  For $N > 4$ except $N = 5$ and $N = 10$, the first peak at $\Lambda_N$ is narrow and not connected to other peaks, indicating HPs are locally caged at their spherical code points\cite{Weeks}.  The width of a peak is proportional to the rattling of a HP within its local cage.  As $\Lambda^c$ increases, the peaks broaden and eventually become connected.   This broadening and overlapping is associated with the degradation of the well-defined structure by increased rattling and sporadic rearranging.  In no case did we find any evidence of new structures emerging.  We note that for the clusters $N = 5$ and $N = 10$ the peaks are not distinct at the smallest $\Lambda^c$ considered.  
For all $N$, if $\Lambda^c >> \Lambda_N$, then the HPs sample uniformly random arrangements on the CP surface and the $n(\theta)$ distribution is a cosine function of $\theta$, truncated to zero when $\theta$ is less than the angular diameter of the HP (e.g. in  Fig.~\ref{fig:Melting}, $N$ = 2, $n(\theta)$ is a truncated cosine function for each $\Lambda^c$).  In Fig.~\ref{fig:Melting} for $N > 2$, insofar as the distributions are far from converged to a cosine function, we observe structure derived from the underlying spherical code solution over the entire range of $\Lambda$ considered

\subsection{Mobility of $N$-clusters} 

We next consider the dynamics of the HPs on the CP surface.  As described in section IIID, we can measure how rapidly the angular displacements of the HPs decorrelate at a given $\Lambda^c$. We call this measure the mobility parameter, $\tau$.   When $\tau = 0$, the HPs are in the first dynamical range; that is, each HP is fully caged and the angular displacement between any two HPs does not decorrelate.  When $\tau > 0$, but small, the HPs are in the second dynamical range. In Figure~\ref{fig:mobility}, the $\tau$ of different clusters are calculated as a function of increasing $\Lambda^c$ relative to the ratio at which the cluster is predicted to assemble, $\Delta \Lambda^c = \Lambda^c - \Lambda_N$.    We observe that the size of the first dynamical range varies widely among clusters.   Noticeably, the $N=6$ and $N=12$ clusters are not measurably mobile until $\Delta \Lambda^c$ is large.   Note that in Fig.~\ref{fig:Melting}, the midpoint $n(\theta)$ data of  both $N$ = 6 and $N$ = 12 still have distinct separated peaks.  In contrast, the $N=11$ cluster becomes mobile at a much lower $\Delta \Lambda^c$ than $N$ = 12 cluster, despite having nearly the same spherical code solution and $\Lambda_{N=11} = \Lambda_{N=12}$.  Comparing the $N$ = 11 and $N$ = 12 distributions in Fig.~\ref{fig:Melting}, at $\Lambda^c$ = 0.9021 the two clusters have nearly identical $n(\theta)$ distributions.  At $\Lambda^c  = 0.9489$, $N=11$ is measurably mobile ($\tau=1.5\cdot10^{-4})$) but still has distinct peaks in $n(\theta)$ that are only slightly softer than that of $N=12$. By $\Lambda^c$ = 1.095 the $n(\theta)$ peaks are noticeably softened for the $N$ = 11 cluster relative to that of the $N$ = 12 cluster and the peaks are connected.  At this ratio, the mobility of the $N=11$ cluster is $\tau=0.19$ while the $N=12$ cluster is just measurably mobile ($\tau = 4\cdot10^{-4}$).   In comparison, the $N$ = 10 cluster is mobile even at the ratio at which it first self-assembles.  The rapid increase of the mobility as a function of increasing $\Lambda^c$ in Fig.~\ref{fig:mobility} is consistent with the connected peaks and the rapid softening of the peak structure for $N$ = 10 in Fig.~\ref{fig:Melting}.   

The fact that HP mobilities for a particular cluster depend upon the cluster's structure is not surprising.  However, it is not obvious that clusters of different sizes should have such variation in the widths of the first dynamical range indicated in Fig.~\ref{fig:mobility}.  There is little correlation, for example, between the mobility of the $N$ cluster and the range of $\Lambda$ over which the $N$ cluster is stable in Fig.~\ref{fig:avha}. We find that the HPs in the $N=5$ and $N=10$ clusters are never fully caged, while the HPs for $N=6$ and $N=12$ are fully caged for a large range of $\Lambda^c$.  The mobilities for $N = 7$, 8, 9, and 11 lie between these extremes.   Note that $N=6$ and $N=12$ clusters have highly symmetrical spherical code point arrangements with octahedral and icosahedral structures, respectively.    Their HP centers define the vertices of Platonic solids with equilateral triangle faces.  For $N$ = 7, 8, 9, 10, and 11, the convex polyhedra defined by the centers of the HP have pentagonal ($N$ = 11), square ($N$ = 8 and 10) or nearly square ($N$ = 7, 9, and 10) faces.  We hypothesize that these non-triangular ``defects'' in the spherical code solutions are responsible for the increased mobility of these clusters by providing locations where the barrier to rearrangement is low.  However, for such small systems, the rearrangements of HPs in a mobile cluster is more appropriately viewed as a rearrangement of the entire cluster rather than a localized rearrangement.

\begin{figure}[h!]
\center
\includegraphics[scale=0.5]{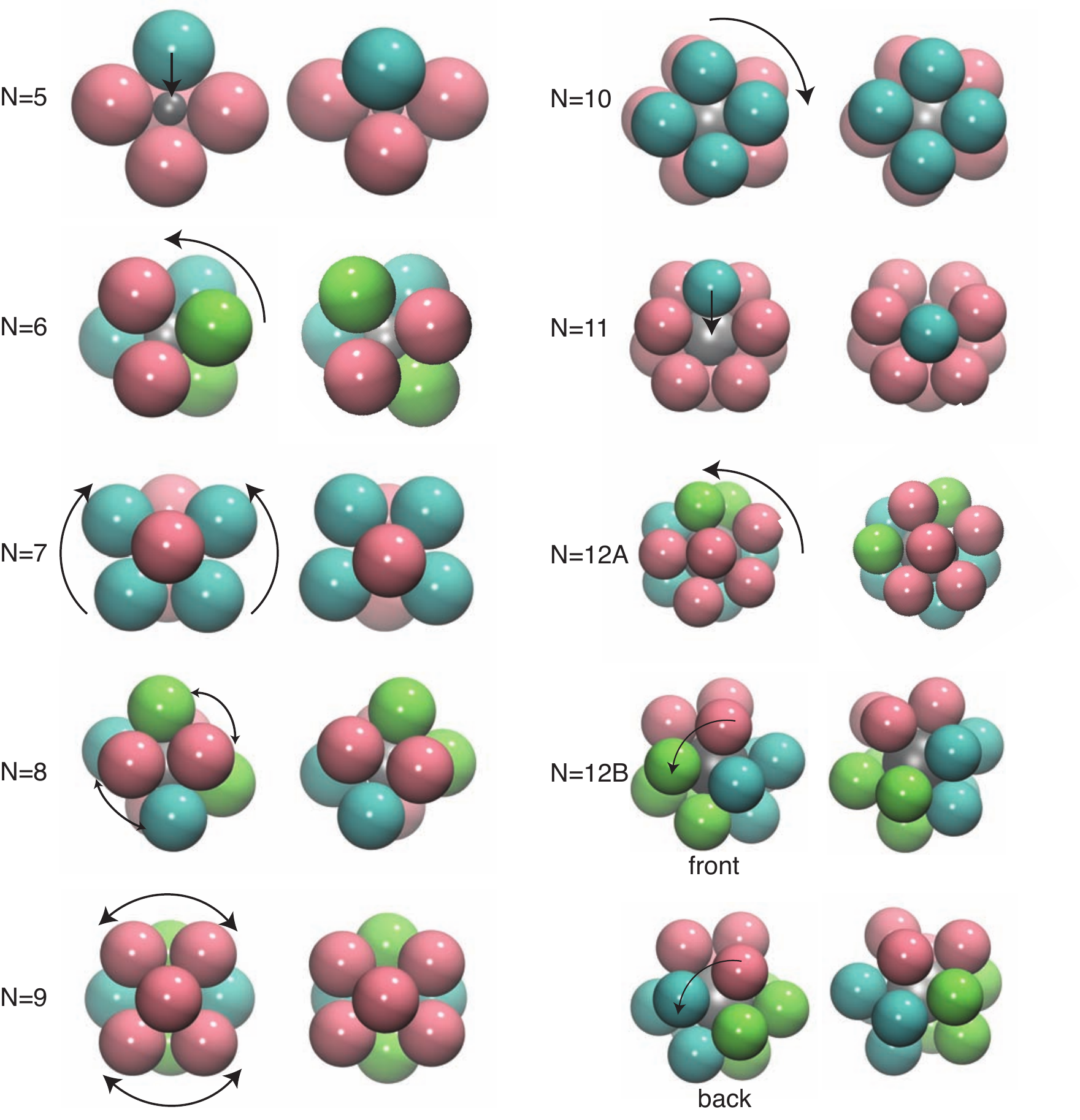}
\caption{(Color) The rearrangements of clusters $N$ = 5-12.  $N$ = 12 has two rearrangements. \label{fig:moves} }
\end{figure}%

As discussed above, in the second dynamical range, the HPs in an $N$ cluster can almost always be mapped to the points of the $N$ spherical code solution, but that mapping does not remain invariant.  We examine the $N$ = 5-12 clusters at the $\Lambda^c$ at which each cluster is first observed to be measurably mobile per Fig.~\ref{fig:mobility} to understand how the HPs in a cluster rearrange. We find (Fig.~\ref{fig:moves}) that the $N$ = 5-11 clusters each have a single unique (discounting reflections or rotations) rearrangement, which permutes the HPs over the spherical code points.  The $N$ = 12 cluster has two unique rearrangements.  Short movies of these rearrangements can be found online in the supplemental material.  For the $N$ = 5 and $N$ = 11 clusters, which have structures equivalent to the $N+1$ spherical code minus a single point, a rearrangement consists of a single HP ``hopping'' a gap to the available $N+1$ point. (The reason the $N$ = 5 is found in this particular configuration is discussed in the next section.)   The clusters $N$ =  6, 10, and 12 exhibit a permutation whereby a ring of HPs rotate relative to the cluster in a manner resembling a twist of a \emph{Rubik's Cube\texttrademark}.   The clusters $N$ = 7, 8, 9, and 12 exhibit a permutation whereby the cluster ``buckles'' into a new permutation of the spherical code points.  We find that the addition of the rearranging action for $N$ = 5-12 is sufficient to make each cluster ergodic.  That is, every possible assignment of each HP to the spherical code points can be explored by the cluster with no inaccessible microstates.  This ergodicity is shown, using group theory, in the supplemental materials.

For clusters $N$ = 6, 7, 8, 9, 10, and 12, the rearranging action is a collective motion of particles in the cluster. Although this entropic contribution is not considered by the free energy calculation in Section \ref{sec:fec}, the free energy calculations compare well with the BD simulations, demonstrating that local rattling is more important than collective modes for some ranges of $\Lambda$.

\subsection{Breaking the degeneracy for $N$ = 5}\label{breaking} \label{magic5}

\subsubsection{BD simulations}
The $N=5$ spherical code has a continuum of solutions ranging from the vertices of a square pyramid to a triangular bipyramid.  For dense $N=5$ clusters at non-zero temperature, we seek the relative likelihood of the cluster adopting particular configurations from the solution continuum. For this, we construct an order parameter that can distinguish between different configurations in our BD simulations.    

All $N$ = 5 spherical code solutions have two points at opposite poles of the central sphere and differ by the positions of the three remaining points on the equator.   The order parameter is constructed by, first, dividing the five HPs into ``pole'' HPs and ``equator'' HPs.  The neighbor distances, or distance between each HP and the four other HPs is measured.  HPs that do not have one neighbor distance that $>$ 1.2 times the distance of the other three neighbor distances are ``equator'' HPs.  Second, the ``equator'' HP that is closest to other ``equator'' HPs or has the minimum summed neighbor distances is selected and and its center is labeled $A$.
   
\begin{figure}[tph]
\includegraphics[scale = 1.0]{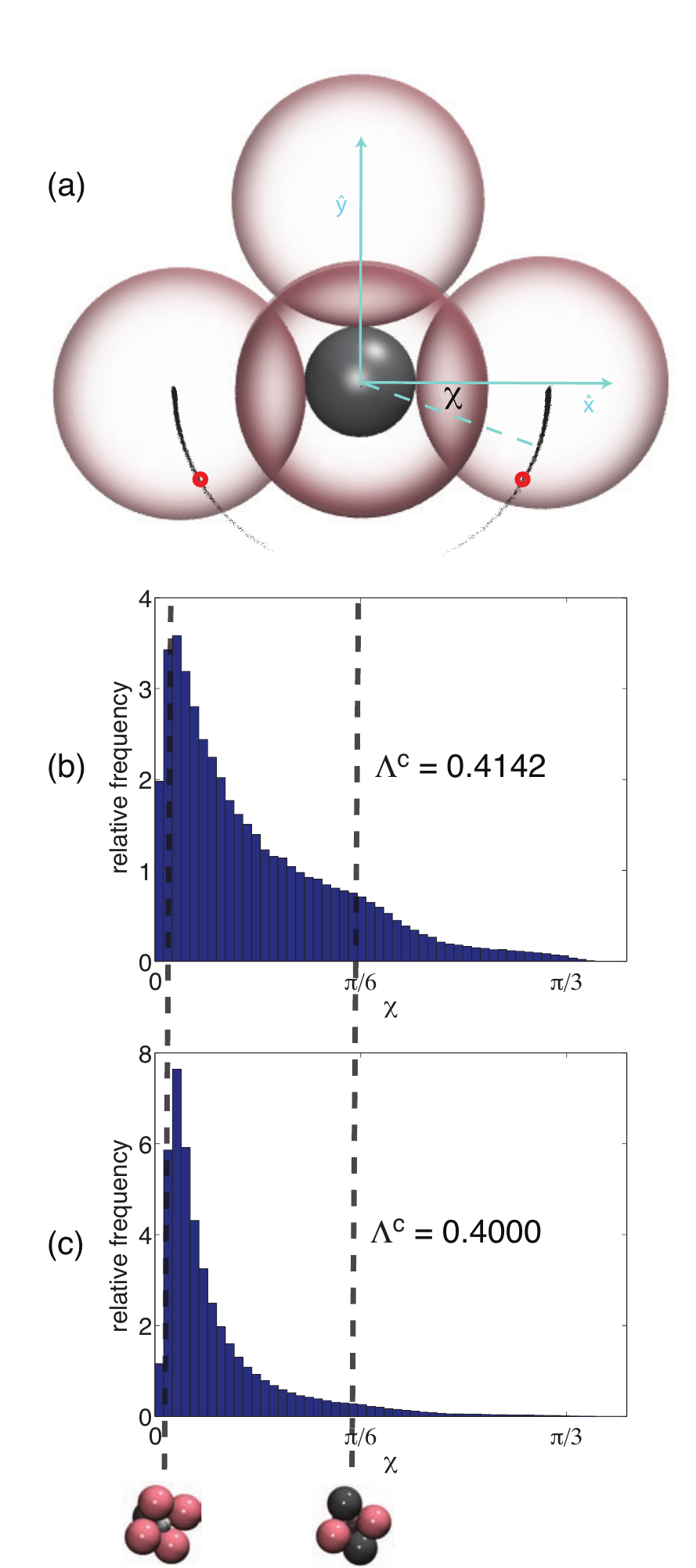}
\caption{(Color) (a) The order parameter $\chi$ is constructed by measuring the angle of the particles on the equator.  Scattered points from a simulation overlay an image of an SP configuration.  Red circles indicate the sphere centers of a TBP configuration.   In (b) and (c) the distribution of $\chi$ sampled in from a BD simulation is shown as a function of the diameter ratio $\Lambda^c$ = 0.4142 and 0.4 respectively.  \label{fig:SBPDist} }
\label{}
\end{figure}

Finally, an angle measurement is constructed in the plane of the equator as follows. The centers of the pair of ``pole'' HPs are labeled $P_1$ and $P2$.  The points $A$, $P_1$, and $P_2$ define a plane $S_1$.  The centers of the two remaining HP are labeled $E_1$ and $E_2$.  The line through $E_1$ and $E_2$ intersects $S_1$ at $E_S$ and $\hat{n}$ is the normal vector to $S_1$.  A plane $S_2$ orthogonal to $S_1$ is constructed from the point $E_S$, $A$, and $A + \hat{n}$.  The coordinates are translated and rotated so that $A$ and $E_S$ are both on the $y$-axis of $S_2$ and $A$ has $x$-$y$ coordinates (0, $r_0$), where $r_0$ is the distance between the center of an HP and the CP.  The origin corresponds to the center of the CP.  The points $E_1$ and $E_2$ are projected to the plane $S_2$ and the angles ($< \pi/2$) to the $x$-axis of $S_2$ is measured.  The order parameter $\chi$ is defined as this angle, sampled twice per configuration.  Each angle pair uniquely specifies a configuration in the solution continuum.  A perfect square pyramid configuration corresponds to two measurements of $\chi=0$ and a perfect triangular bipyramid configuration corresponds to two measurements of $\chi=\pi/6$ ($\approx 0.524$) radians.  Fig.~\ref{fig:SBPDist}a illustrates how the order parameter was constructed, and shows a sampling of the HP positions in the $S_2$ plane from a simulation at $\Lambda^c = 0.4$.  The red circles correspond to the triangular bipyramid positions. 

We performed BD simulations of clusters of 5 HP at  $\Lambda^c=  0.4142$ and $0.400$ with $T^* = 0.02$.  Two histograms are shown of the sampled $\chi$ at the two ratios, 0.4142 and 0.4 in Fig.~\ref{fig:SBPDist}a and \ref{fig:SBPDist}b, respectively.  The figures show that the degenerate continuum of $N=5$ spherical code solutions is broken by the introduction of thermal noise.   Surprisingly, we find that the square pyramid is the preferred structure, even over the more symmetrical triangular bipyramid.   As the cluster is packed tighter, an even stronger preference for the square pyramid configuration over other configurations emerges.   

\subsubsection{Free Energy} 

\begin{figure}[h!]
\includegraphics{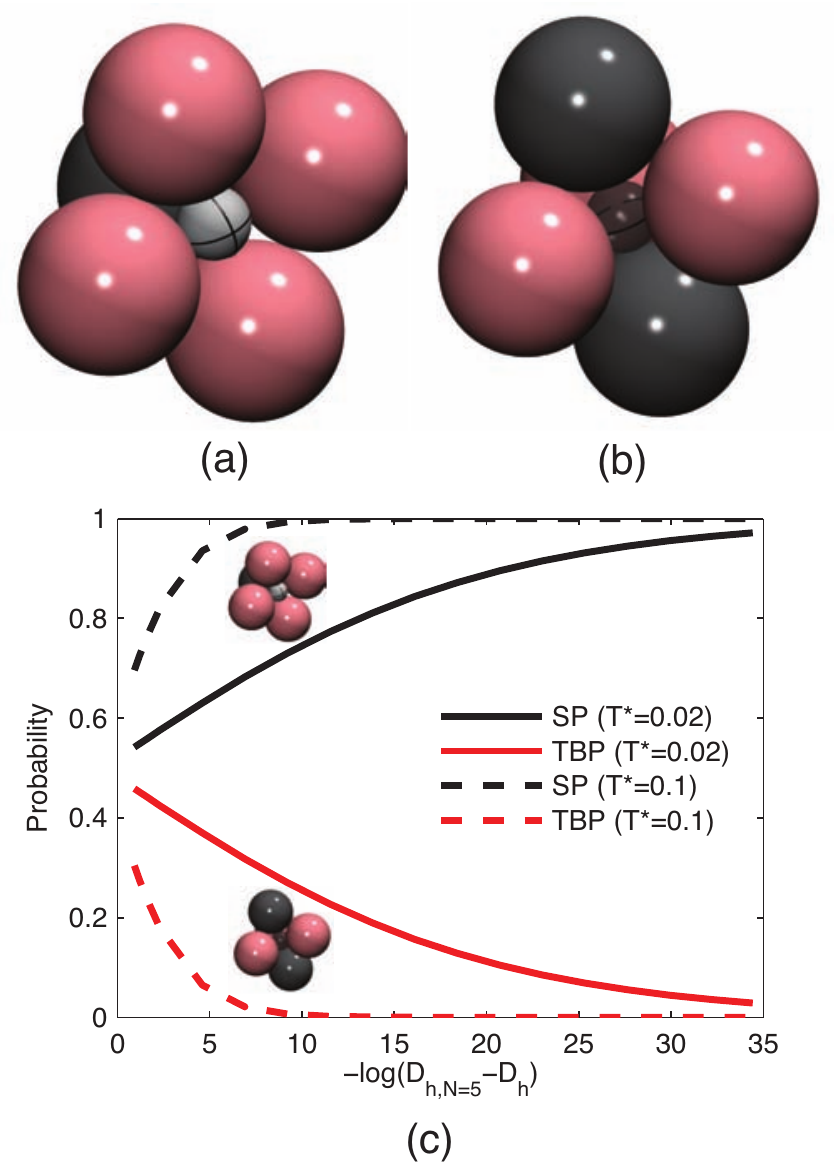}
\caption{(Color) \label{fig:SPandTBP}(Color) (a) The square pyramid (SP) and (b) the triangular bipyramid (TBP) $N$ = 5 spherical codes.  The jammed and unjammed kissing spheres in each configuration are colored dark grey and pink, respectively.   The path that the unjammed spheres can follow is traced on the central sphere.  For (b) the central sphere is transparent so the full path around the equator can be seen.  In the graph at the bottom, at the low temperature, $T^*=0.02$, the preference for the SP (black solid) over the TBP (red solid) is evident as the HP diameter approach the limiting packing diameter.  This preference (black and red dashed lines) is even stronger at high temperature, $T^*=0.1$.}
\end{figure}

To understand the preference for the square pyramid configuration in the BD simulation we use a free energy calculation, which elucidates the role of entropy in breaking the degeneracy.  The more symmetrical triangular bipyramid configuration is used as the reference configuration.

The square pyramid and triangular bipyramid HP clusters are shown in Fig.~\ref{fig:SPandTBP}(a) and Fig.~\ref{fig:SPandTBP}(b).  In Fig.~\ref{fig:SPandTBP}(c), using a free energy calculation, the probability of observing the square pyramid relative to the triangular bipyramid is shown at two temperatures as the HP diameter, $D_h$, approaches the diameter of spheres corresponding to the densest possible packing, $D_{h,N=5}$. For dense clusters, we observe the square pyramid is always the most likely configuration at nonzero temperature.  We find this preference is because the square pyramid has the most vibrational freedom.  In Fig.~\ref{fig:SPandTBP}(a) and Fig.~\ref{fig:SPandTBP}(b), the locally unjammed HP in a cluster at $\Lambda = \Lambda_{N=5}$ are colored pink and  the HP that are locally jammed are colored grey.  In the degenerate continuum of $N=5$ spherical code solutions, only the square pyramid has only one locally jammed HP, and thus, the highest vibrational freedom.

In Fig.~\ref{fig:mobility}, the $N=5$ cluster becomes decreasingly mobile as the cluster is packed tighter, i.e. on decreasing $\Delta \Lambda^c$.  Unlike clusters for other values of $N$, as $\tau \rightarrow 0$, the HPs in the $N=5$ cluster become locally caged for entropic, rather than energetic, reasons.
\section{Discussion}

There are a number of ways the sticky sphere assembly method described above can be extended to create interesting new species of anisotropic particles.   

For example, we can now ponder a more general question.  Given a desired arrangement of points, what HP-CP interactions and HP-HP interactions will result in self-assembly of the arrangement?   The analogous mathematical question was posed by L.L. Whyte in 1952, \emph{``What spherical arrangements [of points] possess extremal properties of any kind?''}\cite{1952}   Ideally, we seek HP-HP interactions and HP-CP interactions that self-assemble repeatable and desirable patterns of HPs on the CP.  

Cohn and Kumar\cite{cohn1} show that all potential energy functions of distance that are completely monotonic, such as inverse power laws, share a subset of universally optimal solution configurations.  If the function is strictly completely monotonic, then the universally optimal solution is also unique.  For points on the surface of a sphere, the only known universally optimal solutions are \cite{cohn1,cohn2} $N$ = 1-4, 6, and 12; that is, a single point, antipodal points, points forming an equilateral triangle on the equator, and tetrahedral, octahedral, and icosahedral arrangement of points. For our purposes, this means certain desired point arrangements (e.g. a ring of 12 points distributed around the equator of a sphere such as modeled in reference \cite{zhangpatchy}) are likely to be inherently difficult to achieve from HP-HP interactions.  Restricting themselves to isotropic pair potentials and identical particles, Cohn and Kumar\cite{cohn3} constructed separate decreasing convex potential energy functions that have cubic (N=8) and dodecahedral (N=20) configurations as their minimum.  Thus references \cite{cohn1, cohn2, cohn3} imply that to assemble certain clusters, it will be necessary to use more complicated HPs with carefully constructed potentials, including non-completely monotonic  or anisotropic interactions.

Using an alternative approach, complex clusters may also be possible by simply adding stages to the assembly process.  For example, if, after the terminal $N$-cluster of Fig.~\ref{makingpatchyparticles} is created, the bath of HPs is replaced by a bath of new HPs coated with the same complementary material as the CP, a second shell of spheres can be added to the first.  The structure of this shell will also depend on the entropy and energy of the cluster at a given temperature.   If the HPs in the second shell preferentially sit in the interstices of the first shell, the polyhedron they form will be the \emph{dual} of the polyhedron of the first shell.  This can make new types of point arrangements possible.  For example, the dual of the octahedron is the cube.  A cubic arrangement of eight points on the surface of a sphere is not found as a minimum among most common spherical surface functions\cite{njas}.   A second shell of HPs that preferentially assemble the dual of the first shell of HPs may be a physically more viable method of assembling a cubic arrangement of spheres without requiring the elaborately constructed HP-HP interaction potential of Cohn and Kumar\cite{cohn3}.

The results presented here may also be used to guide the synthesis of reconfigurable $N=5$ clusters.   As shown in section \ref{magic5} above, a small change in the packing fraction of the $N=5$ cluster introduces a significant change in the structure of the cluster.  Thus, changing the effective diameter of the central particle by a modest amount induces a switch between a relatively isotropic disordered cluster and an anisotropic square pyramidal cluster.

There may also be a correspondence between other mathematical sequences of points distributed on a sphere and terminal cluster assembly problems.  For example in reference \cite{Ting, Ting2}, clusters of cones and spheres were shown to form unique and precisely packed clusters arising from free energy minimization subject to a convexity constraint.  The authors find a packing sequence identical to that obtained from minimizing the second moment of the mass distribution of a cluster of particles constrained to a convex hull. We observe that the packing sequence produced in \cite{Ting} also bears a strong resemblance to the distribution of points on the surface of a  sphere that maximizes the convex hull \cite{njas}.The authors further showed that sequence successfully describes the polyhedral structures formed by colloidal spheres self-assembling on an evaporating droplet\cite{Manoharan25072003,Lauga}.  That work serves another example of the correspondence between mathematical solutions of extremal points on the surface of a sphere and cluster structures obtainable in experiments. 

Another interesting variant to consider is a shaped central particle, as was done in reference \cite{Monica} for the Thomson problem.  Considering the packing of HP around a shaped CP may lead to novel clusters and is a generally unexplored problem.
  
\section{Conclusion}
In this paper we have demonstrated that hard and sticky spheres can self-assemble into terminal $N$-clusters with interesting and, in some cases, unexpected, anisotropies.  These clusters have predictable preferred structures that depend on temperature and sphere diameter ratio.  We find that some clusters exhibit collective particle rearrangements, and these collective modes are unique to a given cluster size. 

If assembled directly from a bath at low temperature, certain cluster sizes (e.g. $N=4,6,12$) form robustly, while other clusters occur only over small ranges with relatively mobile structures (e.g. $N$ = 7,9,10) and still others cannot be formed at all (e.g. $N$ = 5,11).  A ``multi-step'' process that assembles the clusters from a bath at a higher temperature, removes the bath, and lowers the temperature may enable these hard-to-form clusters to be formed robustly as well.  It may even be possible to adjust the effective diameter of the HP or CP as a step in the assembly process. Our free energy calculations and Brownian (molecular) dynamics predictions of cluster structure provide a guide for designing such a process for optimal yield of a desired cluster size with a well-ordered structure.  Clusters fabricated in this way may find use as building blocks for subsequent self-assembly, as templates for manufacturing precisely placed circular patches on the surface of a spherical particle, creating nanocolloidal cages, or fabricating reconfigurable particles.   
 
\subsection{Acknowledgements}  We acknowledge Oleg Gang and Alexei Tkachenko for discussions of related problems.  We acknowledge Daphne Klotsa for her helpful comments on the manuscript.  CLP, MM and SCG were supported by the U.S. Department of Energy, Office of Basic Energy Sciences, Division of Materials Sciences and Engineering, under award 
DE-FG02-02ER46000, the U.S. Department of Energy Computational Science Graduate Fellowship. EJ 
received support from the James S. McDonnell  Foundation 21st Century Science 
Research Award/Studying Complex Systems, grant no. 220020139 and from a National 
Defense Science and Engineering Graduate (NDSEG) Fellowship, 32 CFR 168a. SCG is also supported by the DOD/DDRE under the 
National Security Science \& Engineering Faculty Fellowship award No. N00244-09-1-0062.  Any opinions, findings, and conclusions or recommendations expressed in this publication are those of the author(s) and do not necessarily reflect the views of the DOD/DDRE.   This work was also supported by the Non-Equilibrium Energy Research Center (NERC), an Energy Frontier Research Center funded by the U.S. Department of Energy, Office of Science, Office of Basic Energy Sciences under Award Number DE-SC0000989.

\end{document}